\documentclass{elsart3}

\usepackage{epsfig}

\usepackage{amssymb}
\begin{document}

\begin{frontmatter}



\title{WARP liquid argon detector for dark matter survey}


\author[Pavia]{R. Brunetti}, 
\author[Pavia]{E. Calligarich}, 
\author[Pavia]{M. Cambiaghi}, 
\author[Napoli]{F. Carbonara},
\author[Napoli]{A. Cocco}, 
\author[Pavia]{C. De Vecchi}, 
\author[Pavia]{R. Dolfini}, 
\author[Napoli]{A. Ereditato}, 
\author[Napoli]{G. Fiorillo},
\author[Pavia]{L. Grandi}, 
\author[Napoli]{G. Mangano}, 
\author[Pavia]{A. Menegolli}, 
\author[Pavia]{C. Montanari},
\author[Pavia]{M. Prata},
\author[Pavia]{A. Rappoldi}, 
\author[Pavia]{G.L. Raselli}, 
\author[Pavia]{M. Roncadelli},
\author[Pavia]{M. Rossella}, 
\author[Pavia,enea]{C. Rubbia},
\ead{carlo.rubbia@cern.ch} 
\author[Napoli]{R. Santorelli}, 
\author[Pavia]{C. Vignoli}

\address[Pavia]{INFN and Department of Physics at University of Pavia, Via Bassi 6,  I27100 Pavia (PV), Italy}
\address[Napoli]{INFN and Department of Physics at University of Napoli, Via Cintia, 80126 Napoli (NA), Italy}
\address[enea]{ENEA, Presidenza, Lungotevere Thaon Di Revel 76, 00196 Roma, Italy}
\author[cor]{\textit{Presented by C. Rubbia}}
\corauth[cor]{Spokesman at the 6th UCLA Symposium on \textit{Sources and Detection of Dark Matter and Dark Energy in the Universe}.}

\begin{abstract}
The WARP programme is a graded programme intended to search for cold Dark Matter in the form of \emph{WIMP's}. These particles may produce via weak interactions nuclear recoils in the energy range ${10-100\ keV}$. A cryogenic noble liquid like argon, already used in the realization of very large detector, permits the simultaneous detection of both ionisation and scintillation induced by an interaction, suggesting the possibility of discriminating between nuclear recoils and electrons mediated events. A $2.3\ litres$ two-phase argon detector prototype has been used to perform several tests on the proposed technique. Next step is the construction of a ${100\ litres}$ sensitive volume device with potential sensitivity a factor ${100}$ better than presently existing experiments.
\end{abstract}

\end{frontmatter}

\section{Introduction}
\label{introduction}
\indent There is growing evidence that a large fraction of matter in the Universe is dark and that galaxies are immersed in a dark halo which out-weights the visible component. Elementary particle physics offers an attractive solution in the form of relic, weakly-interacting neutral particles produced shortly after the {\it Big Bang} and pervading cosmic space. Both relativistic particles at the time of the structure formation (\textit{Hot Dark Matter, HDM}) and non relativistic particles (\textit{Cold Dark matter, CDM}) have been considered. \textit{CDM} is highly preferred and one of many possibilities is that it consists of non baryonic weakly interacting massive particles known as \textit{WIMP's} (\textit{Weakly Interacting Massive Particles}). The \textit{Lightest SuperSymmetric Partcle} (\textit{LSP}) of minimal \textit{SUSY} models is one of the most promising \textit{WIMP's} candidate.
\begin{figure}
\begin{center}
\includegraphics[width=7.5cm]{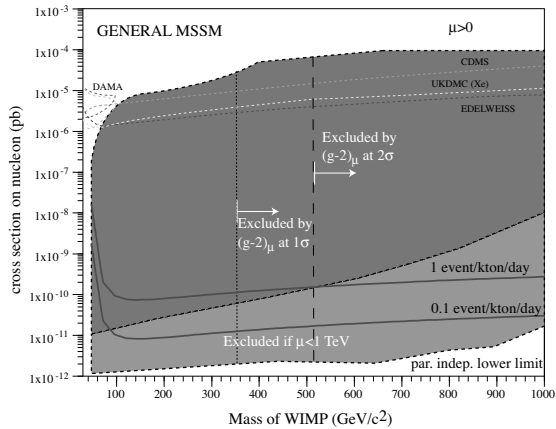}
\caption{\small \textit{Ranges of $\sigma_{p}$  in the general minimal SUSY model as a function of the \textit{WIMP's} mass   for $\mu < 1\ TeV$ which are suggested by collider bounds and $0.1\leq \Omega_{\chi} h^2 \leq 0.2$. The vertical lines show the $m_{\chi}$ upper limit provided by the current discrepancy between the value of the $(g-2)_\mu$ and the Standard Model prediction \cite{ref:kim}. The continued lines show the expected rate for Ar target with a ${30\ keV}$ theshold.}}
\label{fig:1}
\end{center}
\end{figure}
\\
\indent
Hypothetical \textit{WIMP's}, whether \textit{SUSY} or otherwise, could be crossing Earth with ``galactic'' velocity of the order of $\beta \approx 10^{-3}$, presumably with an approximately Maxwellian distribution, eventually truncated at the escape velocity from the Galaxy \cite{ref:lewin}.  A direct detection of  \textit{WIMP's} existence in an underground laboratory, may rely on the observation of the tiny recoils produced in ordinary matter by the elastic scattering with them. The spectrum of such nuclear recoils heavily depends on the choice of the target material \cite{ref:lewin}. If, from one side, the increase in the atomic number improves the well known coherence effect, from the other side it dramatically affects the nuclear form factor $F^{2}(q^{2}$) causing the depletion of the ``gold plated'' events with the largest energy depositions. Evidently the specific signature of these events as recorded by the detector must be such as to isolate them from the much more abundant background due to natural radioactivity and other cosmic ray and solar neutrino events \cite{ref:warp}.\\
\indent
The very small interaction cross section makes \textit{WIMP}-nucleus scattering a very rare event. Such a rate is not easily predicted, since it depends on many variables which are poorly defined. In practice, uncertainties may encompass many orders of magnitude, although the minimal \textit{SUSY} leaves open the optimistic possibility of very significant rates \cite{ref:kim} (see figure \ref{fig:1}). Any new experiment must therefore reach an ultimate sensitivity which is several orders of magnitudes higher than the one of the presently ongoing searches \cite{ref:dama}\cite{ref:cdms}\cite{ref:edelweiss}. To achieve such a goal, both sensitive mass and background discrimination should be as large as possible.\\
\indent
The technology of cryogenic noble liquids (Xe and Ar) permits the detection of both ionisation and scintillation light from multi-ton volumes. Both Liquid Xe and Ar have the potentialities for the kind of sensitive mass required by figure \ref{fig:1}. We believe that the choice of Argon is preferable over Xenon due essentially to the fact that ultra-pure liquid Argon technology is well supported industrially, it has a low cost and it is fully operational. Note that for a realistic energy threshold ($E_R > 30\ keV$) both Argon and Xenon give very similar sensitivities: as explained before the apriori important coherence effect, very rapidly growing with A, is totally absorbed by the steeper form factor. Evidently a much lower threshold, for instance of the order of ${5\ keV}$, will ensure a powerful increase of the rate for Xenon, however in the presence of a substantial background due to coherent neutrino-nucleus scattering of solar neutrinos \cite{ref:warp}.\\
\indent
With the help of a $2.3\ litres$ liquid argon detector chamber exposed to different sources of gamma and neutrons we have studied the proposed discrimination technique and explored the signatures of an hypothetical \textit{WIMP} since fast neutrons scattering elastically on nuclei behave like ``strong interacting'' \textit{WIMP's}, producing nuclear recoils in the energy range of interest.
\section{Tests conducted on a 2.3 litres prototype}
\label{prototype}
\begin{figure}[h!]
\begin{center}
\includegraphics[width=7.5cm]{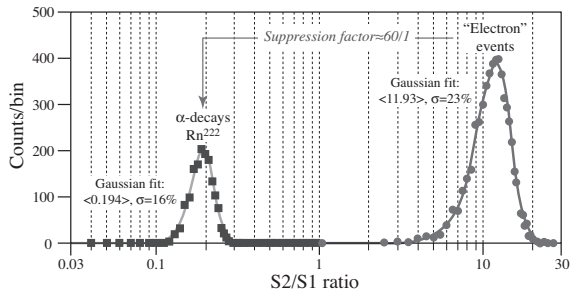}
\caption{\small \textit{Histogram of the distribution of events, plotted as a function $S2/S1$. A clear separation in two peaks is observed. Data are well represented by Gaussian fits of width in agreement with the expected resolutions.} }
\label{fig:2}
\end{center}
\end{figure}

The experimental set up used was a two-phase $2.3\ litres$ drift chamber with a lower liquid volume and an upper region with Argon in the gaseous phase, readout by a single $8$'' cryogenic photomultiplier coated with TPB to wave-shift VUV scintillation photons \cite{ref:warp}. Electrons generated in the liquid are extracted through the liquid-gas boundary with the help of an electric field and detected by the proportional light scintillation generated by accelerating the electrons in a high electric field region. To improve the light collection efficiency from the drift volume, a high performance diffusive reflector layer surrounds the inner volume. A typical light signal associated to an interaction in the liquid volume and recorded by the PMT is then constituted by a prompt primary peak (primary signal $S1$), produced by de-excitations and recombination processes, followed after a drift time (depending on actual location of the interaction) by a secondary peak (secondary signal $S2$), associated to the ionization electrons drifted in the liquid and accelerated in the gas phase. LAr is ultra-purified using the standard procedures developed by the Icarus Collaboration achieving an electronegative impurities concentration $\leq 0.1\ ppb$ ($O_{2}$ equiv.)\cite{ref:warp}.
\\
\indent
Due to their different nature, nuclear recoils and electron mediated process (natural radioactivity, cosmic ray and solar neutrino events) release energy in the medium in different ways, producing different amounts of ionization and excitations. For ion velocities (for ${E_{rec} = 40\ keV}$, ${\beta = 1.4 \times 10^{-3}}$) smaller than the ones of atomic electrons, the classic Bethe-Bloch description of the ionisation process is no longer applicable. At such speeds, single electron collisions are suppressed, and, unlike fast particles, energy losses arise almost exclusively from energy transfers to screened nuclei \cite{ref:lindhard}. The ionization yield (mean energy to produce an ion-electron pair) for a $40\ keV$ argon recoil in argon is about $80\ eV/electron$, about $3$ times higher than the one for minimum ionization particles ($26.4\ eV/electron$) \cite{ref:phipps-hurst}.
\\
\indent 
As a first test the chamber has been exposed to a low activity $^{109}Cd$ source, characterized by a prominent ${X-ray}$ peak in the region $20\div25\ keV$ \cite{ref:warp}. The primary and secondary peaks of the induced signals show a strong correlation and these events are characterized by a ratio between the secondary light produced by electrons extracted into the gas and the primary scintillation produced in the liquid ${S2/S1\gg 1}$. The study of the primary light spectrum (only ${S1}$ considered) has provided, for $E_{drift}=1\ kV/cm$, a photoelectron yield of about $2.35\ phe/keV$.
\\
\indent
The same tests have been conducted with a pulsed and triggered $D-T$ $14\ MeV$ neutron generator \cite{ref:warp}, producing in LAr a nuclear recoil spectrum of mean energy $<E_{rec}>\approx65\ keV$. Differently from gamma-like events the nuclear recoils signals are characterized by ${S2/S1\ll 1}$. For these events, the measured mean number of survived electrons, extracted and multiplied in gas to produce proportional light, is ${<n_{e}>\approx 1.98}$ at $E_{drift}=1\ kV/cm$ (the recombination process is obviously function of the drift field). Such results is in very satisfactory agreement with the one predicted by the so-called Box Model of Thomas and Imel used to modeling the recombination process and providing ${<n_{e}>= 2}$ for a nuclear recoil of energy close to the measured $<E_{rec}>$ \cite{ref:imel}\cite{ref:warp}. On the other side, the study of the primary scintillation spectrum, connected to the nuclear recoil spectrum, has provided a mean photoelectron yield of about $0.66\ phe/keV$ for a mean detected recoil of $65\ keV$(the fit has been executed recalling Lindhard theory)\cite{ref:warp}. The resulting measured quenching factor for argon nuclear recoils in argon is then $f_{N}=0.28$, with  an estimated $10\%$ error, to be compared with the same factors obtained for other scintillators (${f_{N} \approx 0.3}$ and ${f_{N} \approx 0.09}$ for Na and I in NaI, ${f_{N} \approx 0.08}$ and ${f_{N} \approx 0.12}$ for $Ca$ and $F$ in ${CaF_{2}}$ and ${f_{N} \approx 0.16}$ for Xe in Xe). This means that a nuclear recoil of ${65\ keV}$ and an electron of ${18\ keV}$ produce about the same primary signal. Similar results have been obtained with an $^{241}Am-^{9}Be$ neutron source ($2\div 6\ MeV$) \cite{ref:warp}. 
\\
\indent
Figure \ref{fig:2} shows the distribution of ${S2/S1}$ for events recorded in absence of external sources.  As previously explained, both signals, namely the scintillation signal from the liquid and the secondary light produced by the electron emission should be proportional to each other, but with a proportionality factor which depends on the effects of recombination. The strong correlation between $S1$ and $S2$ is clearly evident. Two well separated families of events are visible, one with an experimental ratio $S2/S1$ centred around $11.93$, the other with a much higher depression, $S2/S1= 0.194$. Inspection of single events shows that while the first is due to minimum ionising events, the latter appears when $\alpha$-decays of $^{222}Rn$ are present. Such a signal is actually time dependent, since the decay $^{222}Rn\rightarrow ^{218}Po +\alpha + 5.489\ MeV$ has a half-life of $3.825\ days$. The rate of events follows accurately such $\alpha$-decay dependence. Although in a different energy range the $\alpha$-particles behave like nuclear recoil due to the strong recombination that depletes the proportional light signal. The measured $\alpha$-electron suppression factor of about ${60/1}$ increases of about a factor ${3}$ in the nuclear recoil-electron case at $1\ kV/cm$ . Separation of the two peaks in figure \ref{fig:2} gives an idea of the enormous discrimination power of the technique \cite{ref:warp}.
\section{The proposed 100 litres detector}
\label{100 litres}
\begin{figure}[h!]
\begin{center}
\includegraphics[width=7.5cm]{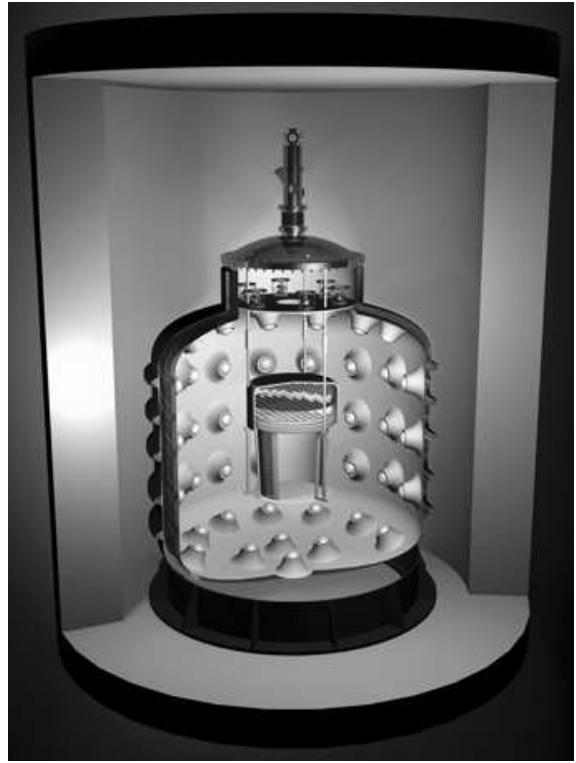}
\caption{\small \textit{Artistic view of the 100 litres prototype.} }
\label{fig:3}
\end{center}
\end{figure}

\indent
On the basis of the comfortable results obtained with the small prototype, a new argon based detector has been proposed \cite{ref:warp}. Its basic scheme foresees a fiducial volume of LAr (about $100\ litres$), tracing the layout of the $2.3\ litres$ chamber, with a uniform electric field drifting ionization electrons towards a liquid to gas interface (see figure \ref{fig:3}). A set of grids with an appropriate voltage arrangement provides then the extraction of ionization electrons from the liquid phase and their acceleration in the gas phase  for the production of the secondary light pulse. A set of photomultipliers placed above the grids sense both the primary scintillation signal in the liquid argon and the delayed secondary pulse in the gas phase. \textit{PMTs} granularity allows reconstruction of event position in the horizontal plane with about ${1\ cm}$ resolution. Position along the drift coordinate is given by the drift time (position reconstructed in 3D). The whole detector has been designed trying to minimize the weight and therefore the amount of materials (and radioactive contamination) to be placed around the inner active volume.
\\
\indent
The detector is completely submersed in a LAr volume that works as an anti-coincidence (\textit{Active VETO}), which is also readout by a set of phototubes, the two volumes are optically separated. The \textit{VETO} region is used to reject the events due to neutrons or other particles penetrating from outside or travelling out from the central part. Dimensions of the outer LAr volume are chosen in such a way that the probability for a neutron to interact in the inner detector without producing a signal in the \textit{VETO} system is negligible. Only events with no signals in it are potential \textit{WIMP's} candidates. \\
\indent
An external shield is added to adequately reduce environmental neutron and gamma background. With ${100\ litres}$ sensitive volume, which corresponds to an active mass of about ${150\ kg}$, thanks to the rejection/identification power of the two-phase technique, we could reach a sensitivity about two orders of magnitude better than the present experimental limit from \textit{CDMS} (or to the presently indicated hint from the \textit{DAMA} experiment). \\ \indent


\end{document}